\documentclass{edbk} 
\usepackage[dvips]{epsfig}
\usepackage{edbkps}

\kluwerbib

\def\be{\begin{equation}}
\def\ee{\end{equation}}
\def\simless{\mathbin{\lower 3pt\hbox
   {$\rlap{\raise 5pt\hbox{$\char'074$}}\mathchar"7218$}}} 
\def\simgreat{\mathbin{\lower 3pt\hbox
   {$\rlap{\raise 5pt\hbox{$\char'076$}}\mathchar"7218$}}}

\begin{document}

\articletitle[Pulsar Kicks]{Physics of Neutron Star Kicks}

\author{Dong Lai}

\affil{Center for Radiophysics and Space Research,
Cornell University, Ithaca, NY 14853, USA}
\email{dong@spacenet.tn.cornell.edu}

\begin{abstract}
It is no longer necessary to ``sell'' the idea of pulsar kicks,
the notion that neutron stars receive a large velocity (a few hundred to a
thousand km~s$^{-1}$) at birth. However, the origin of the kicks remains 
mysterious. We review the physics of different kick mechanisms, including
hydrodynamically driven, neutrino and magnetically driven kicks.  
\end{abstract}

\section{Introduction}

It has long been recognized that neutron stars have space velocities much 
greater than their progenitors'. 
Recent studies of pulsar proper motion give $200-500$~km~s$^{-1}$ as the mean
3D velocity of NSs at birth (e.g., Lyne and Lorimer 1994; Lorimer et al.~1997;
Hansen \& Phinney 1997; Cordes \& Chernoff 1998), with possibly a significant
population having velocities greater than $700$~km~s$^{-1}$. Direct evidence
for pulsar velocities $\simgreat 1000$~km~s$^{-1}$ comes from observations of 
the bow shock produced by the Guitar Nebula pulsar (B2224+65) 
in the interstellar medium (Cordes et al.~1993)
and studies of pulsar-supernova remnant associations (e.g., Frail et al.~1994;
Kaspi 1999). A natural explanation for such high velocities is that
supernova explosions are asymmetric, and provide kicks to nascent neutron
stars. Support for supernova kicks has come from the detection 
of geodetic precession in binary pulsar PSR 1913+16 
(Cordes et al.~1990; 
Kramer 1998; Wex et al.~1999) and the orbital plane
precession in PSR J0045-7319/B-star binary (Kaspi et al.~1996)
and its fast orbital decay (which indicates retrograde rotation of the B star
with respect to the orbit; Lai 1996); These results demonstrate 
that binary break-up (as originally suggested by 
Gott et al. 1970; see Iben \& Tutukov 1996) can not be
solely responsible for the observed pulsar velocities, and that
{\it natal kicks} are required. 
Evolutionary studies of neutron star binary population also
imply the existence of pulsar kicks (e.g.,
Fryer et al.~1998). Finally, there are many direct observations of
nearby supernovae (e.g., 
Wang et al.~1999) and supernova remnants 
which support the notion that supernova explosions are not spherically
symmetric.

Despite decades of theoretical investigations, our understanding 
of the physical mechanisms of core-collapse supernovae remains 
significantly incomplete. While there is a consensus that neutrino heating
of the stalled shock ($0.1-1$~s after bounce) plays an important role
in driving the explosion, it is unclear whether the heating is sufficient
to produce the observed supernova energetics; It is also unclear whether any
convective motion or hydrodynamical instability is central to the explosion
mechanism (e.g., Herant et al.~1994; Burrows et al.~1995; Janka \& M\"uller
1996; Mezzacappa et al.~1998).
The prevalence of neutron star kicks poses a significant mystery, and indicates
that large-scale, global deviation from spherical symmetry is am important
ingredient in any successful theory of core-collapse supernovae. 

In this paper, we concentrate on different classes of 
physical mechanisms for generating 
neutron star kicks (\S\S 2-4), and then briefly discuss the 
astrophysical/observational implications (\S5). 
 
\section{Hydrodynamically Driven Kicks} 

The collapsed stellar core and its surrounding mantle are susceptible
to a variety of hydrodynamical (convective) instabilities 
(e.g., Herant et al.~1994; Burrows et al.~1995; Janka \& M\"uller 1996;
Keil et al.~1996). It is natural to expect that 
the asymmetries in the density, temperature and velocity distributions 
associated with the instabilities can lead to asymmetric matter ejection 
and/or asymmetric neutrino emission. Numerical simulations, however, indicate 
that the local, post-collapse instabilities are not adequate to account 
for kick velocities $\simgreat 100$~km~s$^{-1}$ (Janka \& M\"uller 1994;
Burrows \& Hayes 1996; Janka 1998) --- These simulations were done in 2D,
and it is expected that the flow will be smoother on large scale in 
3D simulations, and the resulting kick velocity will be even smaller. 

Global asymmetric perturbations in presupernova cores are 
required to produce the observed kicks hydrodynamically 
(Goldreich et al.~1996).
Numerical simulations by Burrows \& Hayes (1996) demonstrate that if the
precollapse core is mildly asymmetric, the newly formed neutron star can
receive a kick velocity comparable to the observed values. (In one simulation,
the density of the collapsing core exterior to $0.9M_\odot$ and within $20^o$
of the pole is artificially reduced by $20\%$, and the resulting kick is about
$500$~km~s$^{-1}$.) Asymmetric motion of the exploding material (since the
shock tends to propagate more ``easily'' through the low-density region)
dominates the kick, although there is also contribution (about $10-20\%$)
from asymmetric neutrino emission. The magnitude of kick velocity is 
proportional to the degree of initial asymmetry in the imploding core.
Thus the important question is: What is the origin of the initial asymmetry? 

\begin{figure}
\vskip -0.8cm
\epsfig{file=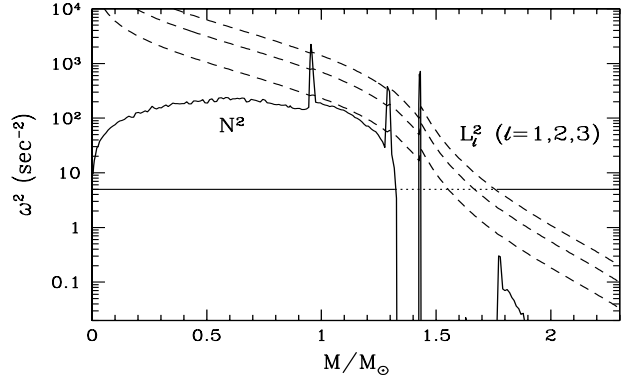, width=10cm, height=8.5cm}
\vskip -3cm
\caption{
 Propagation diagram computed for a $15M_\odot$
presupernova model of Weaver and Woosley (1993). 
The solid curve shows $N^2$, where $N$ is the Brunt-V\"ais\"al\"a
frequency; the dashed curves show $L_l^2$, where $L_l$ is the 
acoustic cutoff frequency, with $l=1,~2,~3$.
The spikes in $N^2$ result from discontinuities in entropy and
composition. The iron core boundary is located at $1.3M_\odot$,
the mass-cut at $1.42M_\odot$.
Convective regions correspond to $N=0$. Gravity modes
(with mode frequency $\omega$) propagate in regions
where $\omega<N$ and $\omega<L_l$, while pressure modes
propagate in regions where $\omega>N$ and $\omega>L_l$.
Note that a g-mode trapped in the core can lose energy
by penetrating the evanescent zones and turning into an
outgoing acoustic wave (see the horizontal line). 
Also note that g-modes with higher $n$ (the
radial order) and $l$ (the angular degree) are better trapped in the
core than those with lower $n$ and $l$.
}\label{fig1}\end{figure}

\smallskip
\noindent
{\bf (i) Presupernova Perturbations}

Goldreich et al.~(1996) suggested that overstable g-mode oscillations in the
presupernova core may provide a natural seed for the initial asymmetry.
These overstable g-modes arise as follows. A few hours prior to core collapse, 
a massive star ($M\simgreat 8M_\odot$) has gone through a successive stages
of nuclear burning, and attained a configuration with a degenerate iron core
overlaid by an ``onion skin'' mantle of lighter elements. The rapidly growing
iron core is encased in and fed by shells of burning silicon and oxygen, and
the entire assemblage is surrounded by a thick convection zone. The nearly 
isothermal core is stably stratified and supports internal gravity
waves. These waves cannot propagate in the unstably stratified convection zone,
hence they are trapped and give rise to core g-modes in which the core
oscillates with respect to the outer parts of the star. The overstability 
of the g-mode is due to the ``$\varepsilon$-mechanism'' with driving provided
by temperature sensitive nuclear burning in Si and O shells surrounding 
the core before it implodes. It is simplest to see this
by considering a $l=1$ mode: If we perturb the core to the right, the 
right-hand-side of the shell will be compressed, resulting in an increase in
temperature; Since the shell nuclear burning rate depends sensitively
on temperature (power-law index $\sim 47$ for Si burning and $\sim 33$ for O
burning), the nuclear burning is greatly enhanced; This generates a large 
local pressure, pushing the core back to the left. 
The result is an oscillating g-mode with increasing amplitude. 

The main damping mechanism comes from the leakage of mode energy. 
The local (WKB) dispersion relation for nonradial waves is 
\be
k_r^2=(\omega^2c_s^2)^{-1}(\omega^2-L_l^2)(\omega^2-N^2),
\ee
where $k_r$ is the radial wavenumber, $L_l=\sqrt{l(l+1)}c_s/r$ ($c_s$ 
is the sound speed) and $N$ are the acoustic cut-off (Lamb) frequency and the 
Brunt-V\"ais\"al\"a frequency, respectively.
Since acoustic waves whose frequencies lie above the acoustic cutoff can
propagate through convective regions, each core g-mode will couple to an
outgoing acoustic wave, which drains energy from the core g-modes (see
Fig.~\ref{fig1}). This leakage of mode energy can be handled with an
outgoing propagation boundary condition in the mode calculation.
Also, neutrino cooling tends to damp the mode.
Since the nuclear energy generation rate depends more sensitively on
temperature than pair neutrino emission (power law index $\sim 9$), cooling is
never comparable to nuclear heating locally. Instead, thermal balance is
mediated by the convective transport of energy from the shells, where the
rate of nuclear energy generation exceeds that of neutrino energy
emission, to the cooler surroundings where the bulk of the neutrino
emission takes place. Calculations (based on the $15M_\odot$ and $25M_\odot$
presupernova models of Weaver \& Woosley 1993) indicate that 
a large number of g-modes are overstable, although for low-order modes
(small $l$ and $n$) the results depend sensitively on the detailed
structure and burning rates of the presupernova models 
(Lai \& Goldreich 2000b, in preparation). 

Our tentative conclusion is that overstable g-modes can potentially grow
to large amplitudes prior to core implosion, although a complete
understanding of the global pre-collapse asymmetries is probably out of reach
at present, given the various uncertainties in the presupernova 
models (see Bazan \& Arnett 1998 for complications due to 
convective shell burning in presupernova stars).  

 
\smallskip
\noindent
{\bf (ii) Amplification of Perturbation During Core Collapse}

Core collapse proceeds in a self-similar fashion, with
the inner core shrinking subsonically and the outer core falling
supersonically at about half free-fall speed
(Goldreich \& Weber 1980; Yahil 1983). The inner
core is stable to non-radial perturbations because of the
significant role played by pressure in its subsonic collapse.
Pressure is less important in the outer region, making
it more susceptible to large scale instability.
A recent stability analysis of Yahil's self-similar collapse solution 
(which is based on Newtonian theory and a polytropic equation of state
$P\propto\rho^\Gamma$, with $\Gamma\sim 1.3$) does not reveal any unstable
global mode before the proto-neutron star forms (Lai 2000).
However,  during the subsequent accretion of the outer core 
(involving $15\%$ of the core mass) and envelope onto the proto-neutron star,
nonspherical perturbations can grow according to $\delta\rho/\rho\propto
r^{-1/2}$ or even $\delta\rho/\rho \propto r^{-1}$ (Lai \& Goldreich 2000). 
The asymmetric density perturbations seeded in the presupernova star,
especially those in the outer region of the iron core, are therefore amplified
(by a factor of 5-10) during collapse. The enhanced asymmetric density
perturbation may lead to asymmetric shock propagation and breakout, which then
give rise to asymmetry in the explosion and a kick velocity to the neutron star
(Goldreich et al.~1996; Burrows \& Hayes 1996).  

\section{Neutrino Driven Kicks}

The second class of kick mechanisms rely on asymmetric neutrino emission
induced by strong magnetic fields. 
The fractional asymmetry $\alpha$ in the radiated neutrino energy required to
generate a kick velocity $v_{\rm kick}$ is $\alpha=Mv_{\rm kick}c/E_{\rm tot}$
($=0.028$ for $v_{\rm kick}=1000$~km~s$^{-1}$, neutron star mass
$M=1.4\,M_\odot$ and total neutrino energy radiated $E_{\rm tot}
=3\times 10^{53}$~erg).

\smallskip
\noindent
{\bf (i) Effect of Parity Violation}

Because weak interaction is parity violating, the neutrino opacities and
emissivities in a magnetized nuclear medium depend asymmetrically on the
directions of neutrino momenta with respect to the magnetic field, and 
this can give rise to asymmetric neutrino emission from the proto-neutron
star. Chugai (1984) (who gave an incorrect expression for the 
electron polarization in the relativistic, degenerate regime) 
and Vilenkin (1995) considered neutrino-electron scattering, but this is less
important than neutrino-nucleon scattering in determining neutrino transport in
proto-neutron stars. Dorofeev et al.~(1985) considered neutrino emission by
Urca processes, but failed to recognize that in the bulk interior of the star
the asymmetry in neutrino emission is cancelled by that associated with
neutrino absorption (Lai \& Qian 1998a). 

Horowitz \& Li (1998) suggested that large asymmetries in the neutrino flux
could result from the cumulative effect of multiple scatterings of neutrinos by
slightly polarized nucleons (see also Janka 1998; Lai \& Qian 1998a). However,
it can be shown that, although the scattering cross-section is asymmetric with
respect to the magnetic field for individual neutrinos, detailed balance
requires that there be no cumulative effect associated with multiple
scatterings in the bulk interior of the star where thermal equilibrium is
maintained to a good approximation (Arras \& Lai 1999a; see also Kusenko et
al.~1998).
For a given neutrino species, there is a drift flux of neutrinos along the
magnetic field in addition to the usual diffusive flux. This
drift flux depends on the deviation of the neutrino distribution function from
thermal equilibrium. Thus asymmetric neutrino flux can be generated 
in the outer region of the proto-neutron star (i.e., above the neutrino-matter
decoupling layer, but below the neutrinosphere) where the neutrino 
distribution deviates significantly from thermal equilibrium. While the
drift flux associated with $\nu_\mu$'s and $\nu_\tau$'s is exactly canceled by
that associated with $\bar\nu_\mu$'s and $\bar\nu_\tau$'s, there is a 
net drift flux due to $\nu_e$'s and $\bar\nu_e$'s. Arras \& Lai (1999b) 
found that the asymmetry parameter for the $\nu_e$-$\bar\nu_e$ flux is
dominated for low energy neutrinos ($\simless 
15$~MeV) by the effect of ground (Landau) state
electrons in the absorption opacity, $\epsilon_{\rm abs}\simeq
0.6B_{15}(E_\nu/1~{\rm MeV})^{-2}$, where 
$B_{15}=B/(10^{15}~{\rm G})$, and for high energy neutrinos by nucleon
polarization ($\sim \mu_mB/T$). Averaging over all neutrino species, 
the total asymmetry in neutrino flux is of order
$\alpha\sim 0.2\epsilon_{\rm abs}$, and the resulting 
kick velocity $v_{\rm kick}\sim 50\,B_{15}$~km~s$^{-1}$.
There is probably a factor of 3 uncertainty in this estimate. To firm up
this estimate requires solving the neutrino transport equations 
in the presence of parity violation for realistic
proto-neutron stars.

\smallskip
\noindent
{\bf (ii) Effect of Asymmetric Field Topology}

A different kick mechanism relies on the asymmetric magnetic field 
distribution in proto-neutron stars (see Bisnovatyi-Kogan 1993;
However, he considered neutron decay, which is not directly relevant for
neutrino emission from proto-neutron stars). Since the cross section for
$\nu_e$ ($\bar\nu_e$) absorption on neutrons (protons) depends on the local
magnetic field strength due to the quantization of energy levels for the $e^-$
($e^+$) produced in the final state, the local neutrino fluxes emerged from
different regions of the stellar surface are different. Calculations
indicate that to generate a kick velocity of $\sim 300$~km~s$^{-1}$ using this 
mechanism alone would require that the difference 
in the field strengths at the two opposite poles of the star 
be at least $10^{16}$~G (Lai \& Qian 1998b). Note that unlike 
the kick due to parity violation [see (i)], this mechanism does not
require the magnetic field to be ordered, i.e., only the magnitude
of the field matters. 

\smallskip
\noindent
{\bf (iii) Exotic Neutrino Physics}

There have also been several interesting ideas on pulsar kicks
which rely on nonstandard neutrino physics. It was
suggested (Kusenko \& Segre 1996) that asymmetric $\nu_\tau$ emission could 
result from the Mikheyev-Smirnov-Wolfenstein flavor transformation
between $\nu_\tau$ and $\nu_e$ inside a magnetized proto-neutron star
because a magnetic field changes the resonance condition for the
flavor transformation. This mechanism requires neutrino mass of order
$100$~eV. Another similar idea (Akhmedov et al.~1997)
relies on both the neutrino mass and the neutrino magnetic 
moment to facilitate the flavor transformation. More detailed 
analysis of neutrino transport (Janka \& Raffelt 1998), however, indicates
that even with favorable neutrino parameters (such as
mass and magnetic moment) for neutrino oscillation, the induced pulsar 
kick is much smaller than previously estimated (i.e.,
$B\gg 10^{15}$~G is required
to obtain $100$~km~s$^{-1}$ kick). 

\section{EM~ Radiation Driven Kicks}

For completeness, we mention the post-explosion
``rocket'' effect due to electromagnetic (EM) radiation from off-centered
magnetic dipole in the pulsar (Harrison \& Tademaru 1975). In this mechanism,
the neutron star velocity comes at the expense of its spin kinetic energy,
which is radiated away asymmetrically via EM braking;
The kick is attained on the timescale of the initial spindown time of the
pulsar (i.e., this is not a ``natal'' kick). The neutron star velocity 
changes according to $M\dot v=\epsilon L/c$, where $L$ is the
EM braking power, and $\epsilon$ is the asymmetry parameter. Typically,
$\epsilon\sim 0.1 (\Omega s/c)(\mu_\phi/\mu_z)$, where $\Omega$ is the spin and
$s$ is the ``off-center'' displacement of the dipole; In fact, there is
theoretical maximum, $\epsilon=0.16$, achieved for $\mu_R=0,~\mu_\phi/\mu_z
=0.63(\Omega s/c)$ (where $\mu_R,\mu_\phi,\mu_z$ are the three cylindrical
components of the dipole). The kick velocity is along the spin axis, and
$v_{\rm kick}\simeq 600\,(\bar\epsilon/0.1)(\nu_0/1~{\rm kHz})^2$~km~s$^{-1}$
(where $\nu_0$ is the initial spin).  
Clearly, Even if the neutron star were born with maximum rotation rate
{\it and} ${\epsilon}$ were maintained at near the maximum value, the kick 
velocity would still be at most a few hundred km~s$^{-1}$. 
Given that most pulsars were born rotating slowly (see Spruit \& Phinney 1998),
we conclude that ``EM rocket'' cannot be the main mechanism for pulsar
kicks.

\section{Discussion}

Statistical studies of pulsar population have revealed no correlation between
$v_{\rm kick}$ and magnetic field strength, or correlation between
the kick direction and the spin axis (e.g., Lorimer et al.~1995; 
Cordes \& Chernoff 1998; Deshpande et al.~1999). Given the large 
systematic uncertainties, the statistical results, by themselves,
cannot reliably constrain any kick mechanism (see Cordes
\& Chernoff 1998). For example, the magnetic field strengths required for the
neutrino-driven mechanisms are $\simgreat 10^{15}$~G, much larger
than the currently inferred dipolar surface fields of typical radio pulsars; 
the internal magnetic fields of neutron stars and their evolution remain
clouded in mystery; and several different mechanisms may contribute to
the observed kick velocities. 
 
It is of interest to note that soft gamma repeaters (``magnetars'':
neutron stars with observed magnetic fields $\simgreat 10^{14}$~G; see 
Thompson \& Duncan 1996) have very high velocities, $\simgreat
1000-2000$~km~s$^{-1}$ (e.g., Kaspi 1999; Marsden et al.~1999). 
Such a high velocity may well require superstrong magnetic fields
($\simgreat 10^{16}$~G) to be present in the proto-neutron stars,  
although hydrodynamical effects remain a viable kick mechanism if 
enough presupernova asymmetry can be generated. It has recently been 
suggested that MHD jets can play an important role in supernovae (Khokhlov et
al.~1999), but the origin of the jets is unknown, nor it is clear why
the two opposite jets are so different (a necessary condition to 
produce a kick).  

\smallskip
I thank my collaborators Phil Arras, Peter Goldreich and 
Yong-Zhong Qian for their important contribution and insight.   
This work is supported by NASA grant NAG 5-8356 and by a fellowship 
from the Alfred P. Sloan foundation.


\begin{chapthebibliography}{1}

\bibitem{Akhmedov97}
Akhmedov, E.~K., Lanza, A., \& Sciama, D.~W. 1997, Phys. Rev. D, 56, 6117

\bibitem{arras99a}
Arras, P., \& Lai, D. 1999a, ApJ, 519, 745 (astro-ph/9806285)

\bibitem{arras99b}
Arras, P., \& Lai, D. 1999b, Phys. Rev. D60, 043001 (astro-ph/9811371)

\bibitem{Bisnovatyi-Kogan93}
Bisnovatyi-Kogan, G.~S. 1993, Astron. Astrophys. Trans., 3, 287

\bibitem{Burrows95}
Burrows, A., Hayes, J., \& Fryxell, B.A. 1995, ApJ, 450, 830

\bibitem{}
Burrows, A., \& Hayes, J. 1996, Phys. Rev. Lett., 76, 352


\bibitem{Chugai84}
Chugai, N.N. 1984, Sov. Astron. Lett., 10, 87

\bibitem{Cordes93}
Cordes, J.M., Romani, R.W., \& Lundgren, S.C. 1993, Nature, 362, 133

\bibitem{Cordes90}
Cordes, J.M., Wasserman, I., \& Blaskiewicz, M. 1990, ApJ, 349, 546

\bibitem{}
Cordes, J.M., \& Chernoff, D.F. 1998, ApJ, 505, 315


\bibitem{Dorofeev85}
Dorofeev, O.F., et al. 1985, Sov. Astron. Lett., 11, 123 

\bibitem{Frail94}
Frail, D.A., Goss, W.M., \& Whiteoak, J.B.Z. 1994, ApJ, 437, 781


\bibitem{Fryer98}
Fryer, C., Burrows, A., \& Benz, W. 1998, ApJ, 498, 333

\bibitem{Goldreich96}
Goldreich, P., Lai, D., \& Sahrling, M. 1996, in
``Unsolved Problems in Astrophysics", ed. J.N. Bahcall and
J.P. Ostriker (Princeton Univ. Press)

\bibitem{}
Hansen, B.M.S., \& Phinney, E.S. 1997, MNRAS, 291, 569

\bibitem{}
Herant, M., et al. 1994, ApJ, 435, 339

\bibitem{Horowitz97b}
Horowitz, C.J., \& Li, G. 1998, Phys. Rev. Lett., 80, 3694 

\bibitem{}
Janka, H.-T., \& M\"uller, E. 1996, A\&A, 306, 167

\bibitem{Janka98}                 
Janka, H.-T. 1998, in ``Neutrino Astrophysics", ed. M. Altmann et al. 

\bibitem{Janka98b}
Janka, H.-T., \& Raffelt, G.G. 1998, Phys. Rev. D59, 023005

\bibitem{Kaspi96}
Kaspi, V.M., et. al. 1996, Nature, 381, 583

\bibitem{kaspi}
Kaspi, V.M. 1999, 
astro-ph/9912284

\bibitem{Keil}
Keil, W., Janka, H.-Th., M\"uller, E. 1996, ApJ, 473, L111

\bibitem{}
Khokhlov, A.M., et al.~1999, astro-ph/9904419

\bibitem{}
Kramer, M. 1998, ApJ, 509, 856

\bibitem{Kusenko96}
Kusenko, A., \& Segr\'e, G. 1996, Phys. Rev. Lett., 77, 4872 (also
astro-ph/9811144)

\bibitem{Kusenko98a}                                       
Kusenko, A., Segr\'e, G., \& Vilenkin, A. astro-ph/9806205


\bibitem{}
Lai, D. 1996, ApJ, 466, L35

\bibitem{}
Lai, D. 2000, ApJ, in press.

\bibitem{}
Lai, D., \& Goldreich, P. 2000, ApJ, in press (astro-ph/9906400)


\bibitem{Lai98a}                         
Lai, D., \& Qian, Y.-Z. 1998a, ApJ, 495, L103 (erratum: 501, L155)

\bibitem{Lai98b}
Lai, D., \& Qian, Y.-Z. 1998b, ApJ, 505, 844

\bibitem{Lorimer97}
Lorimer, D.~R., Bailes, M., \& Harrison, P.~A. 1997, MNRAS, 289, 592

\bibitem{Lyne94}
Lyne, A.G., \& Lorimer, D.R. 1994, Nature, 369, 127

\bibitem{Mazza}
Mazzacappa, A., et al.~1998, ApJ, 495, 911.

\bibitem{Morse95}
Morse, J.A., Winkler, P.F., \& Kirshner, R.P. 1995, AJ, 109, 2104


\bibitem{Spruit}
Spruit, H., \& Phinney, E.S. 1998, Nature, 393, 139
 

\bibitem{Thompson96}
Thompson, C., \& Duncan, R.~C. 1996, ApJ, 473, 322


\bibitem{Vilenkin95}
Vilenkin, A. 1995, ApJ, 451, 700


\bibitem{wang}
Wang, L., Howell, D.A., H\"oflich, P., \& Wheeler, J.C. 1999, 
astro-ph/9912033

\bibitem{wex}
Weaver, T.A., \& Woosley, S.E. 1993, Phys. Rep., 227, 65

\bibitem{wex}
Wex, N., Kalogera, V., \& Kramer, M. 1999, astro-ph/9905331.

\end{chapthebibliography}
\end{document}